# TRIPLET-CONSTRAINT TRANSFORMER WITH MULTI-SCALE REFINEMENT FOR DOSE PREDICTION IN RADIOTHERAPY

*Lu Wen*[1,†], *Qihun Zhang*[1,†], *Zhenghao Feng*[1], *Yuanyuan Xu*[1], *Xiao Chen*[1], *Jiliu Zhou*[1], *Yan Wang*[1,*]

[1]College of Computer Science, Sichuan University, China.

## ABSTRACT

Radiotherapy is a primary treatment for cancers with the aim of applying sufficient radiation dose to the planning target volume (PTV) while minimizing dose hazards to the organs at risk (OARs). Convolutional neural networks (CNNs) have automated the radiotherapy plan-making by predicting the dose maps. However, current CNN-based methods ignore the remarkable dose difference in the dose map, i.e., high dose value in the interior PTV while low value in the exterior PTV, leading to a suboptimal prediction. In this paper, we propose a triplet-constraint transformer (TCtrans) with multi-scale refinement to predict the high-quality dose distribution. Concretely, a novel PTV-guided triplet constraint is designed to refine dose feature representations in the interior and exterior PTV by utilizing the explicit geometry of PTV. Furthermore, we introduce a multi-scale refinement (MSR) module to effectively fulfill the triplet constraint in different decoding layers with multiple scales. Besides, a transformer encoder is devised to learn the important global dosimetric knowledge. Experiments on a clinical cervical cancer dataset demonstrate the superiority of our method.

## 1. INTRODUCTION

Radiotherapy stands as a fundamental treatment approach for cancer patients. In clinic, an acceptable radiotherapy plan requires delivering sufficient radiation dose to the planning target volume (PTV) while mitigating dose hazards to the organs at risk (OARs) as much as possible [1]. To achieve this, the dosimetrists have to involve several trial-and-error steps to iteratively adjust the radiotherapy plan, which is both labor-intensive and time-consuming. Therefore, developing an automatic method to predict the dose distribution for cancer patients is of vital importance to enhance the plan-making efficiency.

Recently, inspired by the powerful ability of convolutional neural networks (CNNs) for automatic feature extraction, several CNN-based methods have been researched to perform an automatic prediction of dose distribution. Their common pipeline is to take the CT images along with segmentation masks of PTV and OARs as input, and directly output the corresponding dose maps [2-12]. For example, Nguyen *et al*. [2] first applied the conventional 2D UNet for predicting the dose maps for prostate cancer patients. Furthermore, Kearney *et al*. [3] proposed DoseNet, a fully connected neural network for prostate dose prediction. Besides, Song *et al*. [4] utilized another classical encoder-decoder network, i.e., DeepLab V3+, to accelerate the radiotherapy planning of rectum cancer. Thanks to the atrous spatial pyramid pooling, DeepLabV3+ could extract more context information from different scales. Treating dose prediction as an image generation task, Mahmood *et al*. [6] employed the generative adversarial network (GAN) with adversarial training to predict dose maps, thus reaching a higher prediction accuracy.

Despite their promising results, these CNN-based methods still have the following limitations. First, they neglect the remarkable dose difference inside the dose map, i.e., high dose value in the interior PTV while low value in the exterior PTV (also coincided with the clinical requirements of the radiotherapy plan). Such critical dose difference can be explicitly indicated by the PTV segmentation mask, which has not been further utilized by current methods. Besides, the intrinsic locality of convolutional operation which only gathers local information from the neighborhood pixels, is insufficient to capture the global dosimetric information for the dose prediction task.

In this paper, to address the above-mentioned limitations, we propose a triplet-constraint transformer with multi-scale refinement, named TCtrans, to automatically predict the dose distribution in radiotherapy. The main contributions of this paper can be summarized as follows: (1) To consider the important dose difference inside the dose map, we propose a novel PTV-guided triplet constraint to refine dose feature representations in the interior and exterior PTV by utilizing the explicit geometry of PTV, thus improving the quality of predicted dose prediction. (2) We devise a multi-scale refinement (MSR) module for assisting the triplet constraint to gradually refine the prediction in the multi-scale decoding

---

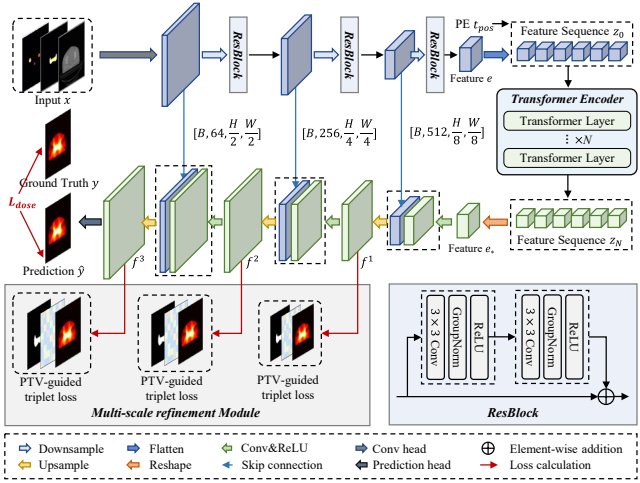

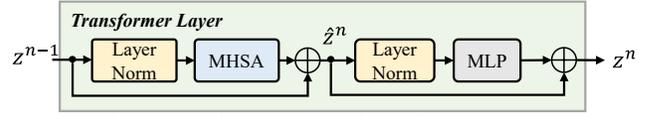

**Fig. 2.** Detailed structure of transformer layer.

**Fig. 1.** Overview of the proposed TCtrans.

stages. (3) Inspired by the powerful ability of long-range sequence modeling of the transformer [13], we embed a transformer encoder into the CNN backbone to learn the global dosimetric information. Extensive experiments conducted on a clinical dataset with 42 cervical cancer patients have demonstrated the superior performance of our proposed method.

## 2. METHODOLOGY

The overview of our proposed TCtrans is illustrated in Fig. 1 which contains three critical parts: (1) a CNN encoder to learn the compact feature representations from the anatomy images, (2) a transformer encoder to further capture the global dependencies among the input sequences from the CNN encoder, (3) a CNN decoder to restore the feature representations processed by the transformer to the final predicted dose map, and (4) a multi-scale refinement module equipped with the novel PTV-guided triplet constraint for further refining the dose feature inside the CNN decoder. More details are described in the following subsections.

### 2.1. Architecture

***CNN encoder***: The cancer patient's image pair can be denoted as $\{x, y\}$ where $x=\{x_{CT}, x_{PTV}, x_{OARs}\}$ is the CT image along with segmentation masks of both PTV and OARs while $y$ is the corresponding dose map. Fed with anatomy image $x$, the CNN encoder outputs its compact feature representation $e$. Concretely, our CNN encoder employs three encoding layers. Each layer is constructed by a residual block which has two $3 \times 3$ convolutional layers accompanied by a Group Normalization and ReLU activation function.

***Transformer encoder***: Considering the intrinsic locality of CNN, a transformer encoder is embedded to learn the global dosimetric knowledge. The feature from the CNN encoder, i.e., $e$, is flattened to $M$ feature tokens and then added with a position embedding (PE), i.e., $t_{pos}$, to maintain their position information, thus gaining a feature sequence $z_0$ as below:
$$z_0 = [t_1; t_2; \ldots; t_M] + t_{pos}. \quad (1)$$
$z_0$ is treated as the input of the transformer encoder with $N$ identical transformers. The detailed structure of transformer layer is displayed in Fig. 2 where each transformer layer is equipped with a multi-head self-attention (MHSA) [13] to explicitly model the interactions among all feature tokens. Denoting the input of the $n$-th transformer layer as $z_{n-1}$ ($z_0$ for the first layer), its corresponding output $z_n$ can gained through the following formulation:
$$z_n = MLP(LN(\hat{z}^n)) + \hat{z}^n, \quad (2)$$
$$\hat{z}^n = MHSA(LN(z^{n-1})) + z^{n-1}, \quad (3)$$
where $\hat{z}^n$ is an intermediate token sequence while $MLP(\cdot)$ and $LN(\cdot)$ respectively represent the multilayer perceptron and layer normalization. Later, the output of the transformer encoder, i.e., $z_N$, is reshaped to a feature map $e_*$ which is then fed into the CNN decoder for further processing.

***CNN decoder:*** Consistent with the CNN encoder, our CNN decoder also contains three decoding layers which gradually up-samples $e_*$ and finally outputs the predicted dose distribution $\hat{y}$. Besides, skip connections are maintained for multi-level feature reuse and aggregation between the encoder and decoder.

### 2.2. PTV-guided triplet constraint with multi-scale refinement

***PTV-guided triplet constraint***: Considering the notable dose difference inside the dose map (i.e., high dose value in the interior PTV while low value in the exterior PTV), we propose a novel PTV-guided triplet constraint to further differentiate the dose feature representations in the interior and exterior PTV, motivated by current metric learning [14]. The illustration of our PTV-guided triplet constraint is shown in Fig. 3. Specifically, we first spit the PTV mask into $K$ patches with a size of $S \times S$ and gain $I$ margin patches according to the PTV boundary. The pixels of $i$-th margin patch can be grouped into triplets: the central pixel as the *anchor* (located inside the PTV for example), these in the interior PTV as *positives*, and these in the exterior PTV as *negatives*. Based on the location of these triplets, the dose feature from the CNN decoder layer (i.e., $f$) can be also separated as the set of positive and negative pixels of the anchor $p_i$, denoting as $p_i^+$ and $p_i^-$. So, the anchor-positive distance $d_i^+$ and anchor-negative distance $d_i^-$ can be expressed as a mean Euclidean distance of dose feature:

$$d_i^+ = \frac{1}{|p_i^+|} \sum_{j \in p_i^+} \sqrt{(f_i - f_j)^2}, \quad (4)$$

$$d_i^- = \frac{1}{|p_i^-|} \sum_{j \in p_i^-} \sqrt{(f_i - f_j)^2}. \quad (5)$$

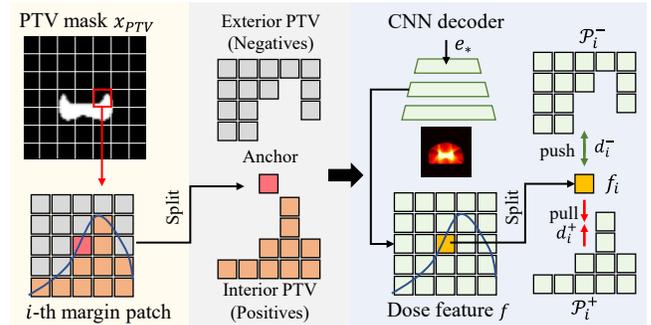

**Fig. 3.** Illustration of the PTV-guided triplet constraint.

To reach the goal of differentiating the dose feature representations in the interior and exterior PTV, we just need to decrease $d_i^+$ while increasing $d_i^-$ through an inner-patch triplet loss:

$$L_i = max(0, d_i^+ + m - d_i^-), \quad (6)$$

where $m$ is a threshold to control the smallest distance between $d_i^+$ and $d_i^-$.

Finally, the proposed PTV-guided triplet constraint loss $L_{tp}$ calculates the patch-wise triplet loss of every margin patch, which can be summarized as:

$$L_{tp} = \frac{\sum_{i \in [1,I]} L_i}{S \times S}. \quad (7)$$

In this manner, the CNN decoding layer can generate more discriminative dose features near the PTV boundary and gain the dose map with the desired difference in dose distribution.

**Multi-scale refinement module**: Instead of directly using the PTV-guided triplet constraint in the last decoding layer of the CNN decoder, we introduce a multi-scale refinement (MSR) module to gradually constrain the dose features. Concretely, for the $R$ decoding layers ($R$ denotes the total number of layers), their corresponding dose feature $f^r$ ($r \in [1, R]$) with different scales is respectively refine by the PTV-guided triplet loss $L_{tp}^r$ and then aggregated by the following equation:

$$L_{mtp} = \sum_{r \in [1,R]} L_{tp}^r. \quad (8)$$

The MSR module utilizes the rectified dose feature in the deeper layer as the input of the shallower one, thus providing multi-scale refinement of the dose feature.

### 2.3. Objective functions

The objective function of the whole network contains two parts: (1) the above-mentioned muti-scale PTV-guided triplet loss $L_{mtp}$, and (2) the commonly used dose prediction loss $L_{dose}$ defined as the L1 distance between prediction $\hat{y}$ of TCtrans and the ground truth $y$:

$$L_{dose} = \|y - \hat{y}\|_1. \quad (9)$$

Therefore, the total loss function can be formulated as:

$$L = L_{dose} + \omega L_{mtp}, \quad (10)$$

where $\omega$ is a hyper-parameter to balance the two loss terms.

## 3. EXPERIMENTS AND RESULTS

### 3.1. Dataset and implementation details

***Dataset***: We measure the performance of our model on an in-house cervical cancer dataset with 42 patients who underwent the volumetric modulated arc therapy (VMAT) treatment at West China Hospital. For every patient, the CT image, PTV segmentation, five OAR segmentations, and clinically planned dose distribution are collected. Five OARs are small intestine (Small), Femoralhead R (FR), Femoralhead L (FL), bladder (BLA), and rectum (REC). We randomly select 28 patients for model training, 2 patients for validation, and the remaining 12 patients for testing. All the 3D volumes are preliminarily sliced into 2D images with a size of $512 \times 512$. In this way, we finally obtain 4801, 342, and 2024 slices for model training, validation, and testing.

***Implementation details***: We fulfill the proposed network in the PyTorch framework. All experiments are conducted through one NVIDIA RTX 3090 GPU with 24GB memory and a batch size of 12 with the SGD optimizer. The number of transformer layers $N$ and patch size $S$ are set as 12 and 5. Following [14], the threshold in Eq. (6) is set as 0.3. The hyper-parameter $\omega$ in Eq. (10) is empirically set as 0.01. We train the whole model for 300 epochs where the initial learning rate is set to 1e-4 and updated with a poly scheduler.

### 3.2. Evaluation metrics

To comprehensively evaluate the prediction performance, we employ several metrics. In clinical practice, $D_x$ denotes the minimum absorbed dose covering $x\%$ volume and we choose $D_{98}$, $D_{95}$, and $D_{mean}$ (mean absorbed dose) as the metrics. Then, we calculate the average absolute prediction errors (Δ) of them to directly quantify the disparity between the prediction and ground truth. Following [11], heterogeneity index ($HI$) is involved to quantify the dose heterogeneity in the predicted target volume. Then, we conduct the paired t-test between our method and the state-of-the-art (SOTA) methods and calculate the corresponding p-values on all metrics. A p-value less than 0.05 (marked by "*") means the enhancement of our proposed is of statistical significance. More intuitively, dose volume histogram (DVH) [15] serves as another metric for assessing the prediction performance. A smaller disparity between the DVH curves of the prediction and ground truth signifies a higher prediction precision.

**Table. 1** Quantitative results of SOTA methods on PTV in terms of $HI$, $|\Delta D_{98}|$, $|\Delta D_{95}|$, and $|\Delta D_{mean}|$, respectively. The best results are in **bold** while the second-best one is underlined.

| Methods | HI | $|\Delta D_{98}|$(Gy) | $|\Delta D_{95}|$(Gy) | $|\Delta D_{mean}|$(Gy) |
|---|---|---|---|---|
| UNet | 0.078(3.22E-4)* | 1.31(1.51)* | 1.21(1.40)* | 1.12(1.42)* |
| DoseNet | **0.022(3.75E-5)** | 1.26(1.15)* | 1.10(0.99) | 0.97(0.97)* |
| DeepLab V3+ | 0.068(8.54E-5)* | 1.31(1.58)* | 1.09(1.36) | 0.81(1.10) |
| GAN | 0.091(1.18E-4)* | 1.14(1.31)* | 1.02(1.19) | 0.99(1.17)* |
| PRUNet | 0.074(1.69E-4)* | 1.31(1.54)* | 1.12(1.31)* | 0.96(1.25)* |
| Proposed | 0.059(4.96E-5) | **1.09(0.84)** | **0.95(0.71)** | **0.79(0.64)** |

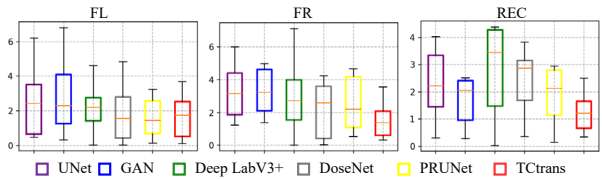

**Fig. 4.** Box plots of SOTA methods on OARs (i.e., FL, FR, and REC) in terms of $\Delta|D_{mean}|$.

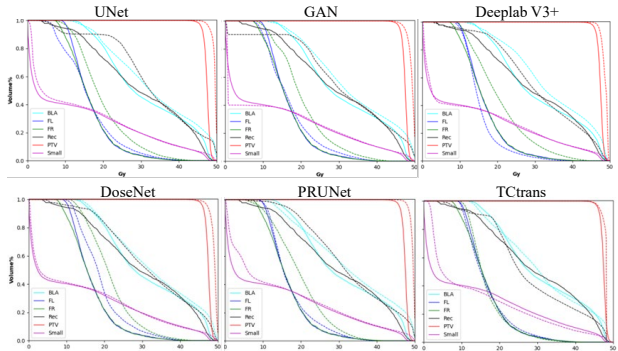

**Fig. 5.** DVH curves of the predicted plan (dashed line) and the manually optimized plan (solid line) for comparison with SOTA methods.

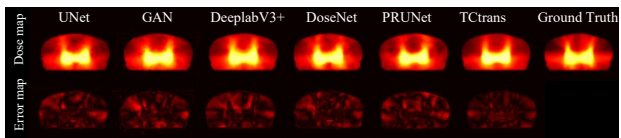

**Fig. 6.** Visual comparisons with SOTA methods.

### 3.3. Comparison with the state-of-the-art methods

To demonstrate the superiority of the proposed TCtrans, we compare it with five SOTA methods in dose prediction, including UNet [2], DoseNet [3], DeepLab V3+ [4], GAN [6], and PRUNet [16]. The quantitative results are summarized in Table 1 where our proposed TCtrans achieves the best overall prediction accuracy. Specifically, UNet shows the poorest performance. Though the DoseNet gains a better performance in *HI*, i.e., 0.022, our method further decreases $|\Delta D_{mean}|$ and $|\Delta D_{98}|$ by 0.18Gy and 0.17Gy, respectively. With the help of adversarial training, GAN performs well on $|\Delta D_{98}|$ and $|\Delta D_{95}|$, but the proposed still drops them from 1.14 to 1.09, and from 1.02 to 0.95, respectively.

For OARs, we also display a box plot on OARs (i.e., FL, FR, and REC) in terms of $\Delta|D_{mean}|$ in Fig. 4. As seen, the proposed still maintains its best performance, especially for FR and REC. In addition, as illustrated in Fig. 5, the DVH curves of the proposed best match the ground truth with the smallest prediction disparity. Furthermore, as seen in the visual comparisons in Fig. 6, our method gains the darkest error map which also validates its superior performance.

### 3.4. Ablation Study

To study the contribution of key components in TCtrans, we perform multiple ablation experiments in an incremental way. The experimental arrangements can be summarized as (A) CNN backbone (Baseline), (B) Baseline + transformer encoder (Baseline + Trans), (C) Baseline + Trans + PTV-guided triplet loss simply applied to the final prediction and the ground truth (Baseline + Trans + TL), and (D) Baseline + Trans + TL + multi-scale refinement module (Baseline + Trans + TL + MSR, proposed). The Quantitative results are listed in Table 2. As seen, compared (B) and (C), the PTV-guided triplet loss reduces the *HI* from 0.067 to 0.060, and $|\Delta D_{95}|$ from 1.05 to 0.99, indicating its efficiency in enhancing the accuracy. Besides, compared (C) to (D), with the help of the MSR module, (D) further drops $|\Delta D_{95}|$ by 0.04, $|\Delta D_{mean}|$ by 0.02. Overall, with the progressive addition of key components, the prediction errors are reduced among all the metrics, thus validating their respective contributions.

**Table. 2** Quantitative results of ablation models in terms of *HI*, $|\Delta D_{98}|$, $|\Delta D_{95}|$, and $|\Delta D_{mean}|$, respectively.

| Methods | HI | $|\Delta D_{98}|$(Gy) | $|\Delta D_{95}|$(Gy) | $|\Delta D_{mean}|$(Gy) |
|---|---|---|---|---|
| (A) | 0.081(3.22E-4)* | 1.15(0.82)* | 1.12(0.71)* | 0.82(0.60)* |
| (B) | 0.067(8.09E-5) | 1.10(0.95) | 1.05(0.96)* | 0.82(0.63)* |
| (C) | 0.060(4.70E-5) | 1.10(0.96) | 0.99(0.70)* | 0.81(0.64) |
| (D) | **0.059(4.96E-5)** | **1.09(0.84)** | **0.95(0.71)** | **0.79(0.64)** |

### 4. CONCLUSION

In this paper, we propose TCtrans, a triplet-constraint transformer with multi-scale refinement, to automatically predict the radiotherapy dose distribution for cervical cancer patients. Considering the important dose difference inside the dose map, TCtrans uses a novel PTV-guided triplet constraint to refine dose feature representations through employing the explicit geometry of PTV. Furthermore, a multi-scale refinement module is designed to help the triplet constraint progressively refine the dose feature representations in the multi-scale decoding layers. Besides, a transformer encoder is introduced to our model to provide a powerful learning of global dosimetric information. Experimental results on the clinical dataset with 42 cervical cancer patients have demonstrated the superior performance of our method.

### 5. COMPLIANCE WITH ETHICAL STANDARDS

## 6. ACKNOWLEDGMENTS

This work is supported by National Natural Science Foundation of China (NSFC 62371325, 62071314), Sichuan Science and Technology Program 2023YFG0263, 2023YFG0025, 2023NSFSC0497, and Opening Foundation of Agile and Intelligent Computing Key Laboratory of Sichuan Province.